# Insights on the interaction of calcein with calcium carbonate and its implications in biomineralization studies

Giulia Magnabosco [a,*], Iryna Polishchuk [b], Jonathan Erez [c], Simona Fermani [a], Boaz Pokroy [b], Giuseppe Falini [a,*]

The effects of calcein, a marker commonly used to assess mineral growth in calcifying organism, on calcite and aragonite structure have been investigated. Calcein is entrapped within calcite and aragonite and modifies the shape and morphology of both polymorphs. Moreover, in the presence of $Mg^{2+}$, it inhibits aragonite formation in favor of magnesium calcite.

Studying and understanding of the biomineralization processes employ various methods to mark the growth of the inorganic components of the organisms, allowing researchers to correlate a particular region of the skeleton to the instant of its deposition. A common method is to use fluorescent molecules to label a particular stage of the deposition process.[1] Calcein, a strongly fluorescent organic dye synthetized via modification of fluorescein with ethylenediaminetetraacetic acid analogues,[2] has been widely used to study the growth of calcified skeletons because it coordinates calcium ions and is entrapped within the growing inorganic matrix.[3-5] Calcein has been successfully applied to mark bone growth [6] as well as shell and coral skeleton formation.[1,7] However, even if the transport of this dye inside living organisms was deeply characterized, [3,8] no information on the effect of calcein on the morphology and polymorphism of calcium carbonate crystals is readily available.

The effect of soluble additives, both synthetic and biogenic, on calcium carbonate crystals has been extensively studied to understand biomineralization[9-13] and to fabricate novel advanced materials[14-20]. Based on these studies, models on the mechanisms of interaction between various additives and the growing mineral have been proposed. Interaction and entrapment of calcein within the crystal structure of calcium carbonates is known to occur due its chemical structure.[21,22]

In this work, we examined *in vitro* the effect of calcein on the growth of calcite and aragonite crystals, the two main polymorphs of $CaCO_3$ found in living organisms, using the calcein concentration in the range usually adopted for *in vivo* labelling.

For this goal, we used the vapor diffusion method, consisting in the diffusion of $NH_{3(g)}$ and $CO_{2(g)}$ obtained from the decomposition of $(NH_4)_2CO_{3(s)}$, into a $Ca^{2+}$ solution containing the dye. This method is relevant for biomineralization process due to the slow increase of the concentration of $CO_3^{2-}$ ions in the crystallization solution. It exploits carbonate speciation in water resembling the process suggested to occur in living systems, in which carbonic anhydrase controls the supply of carbonate to the calcification site. [23]

In our experimental setting, calcite precipitated when solely $Ca^{2+}$ was present in the crystallization solution, while aragonite was the main polymorph obtained when $Mg^{2+}$ is co-present with a $Mg^{2+}/Ca^{2+}$ molar ratio equal to 4. [24,25]

No detectable inhibition or promotion of precipitation due to the presence of calcein was observed, as evaluated by the measure of total $Ca^{2+}$ deposited in each experiment (see table SI1).

To evaluate the quantity of calcein associated to the inorganic phase, the crystals were dissolved in an acidic buffer and the UV-Vis absorption spectrum of calcein was measured. The high molar absorption coefficient of calcein allows an accurate quantification of the dye in solution even for very low concentrations, making this approach more accurate than other techniques (e.g. thermogravimetric analysis).[26]

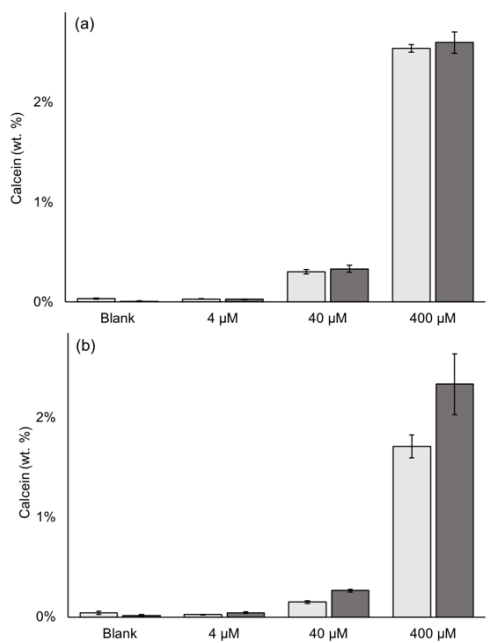

**Figure 1.** Calcein content (wt.%) measured in (a) calcite and (b) aragonite precipitated in the presence of the different dye concentration examined (Table SI1). The data on bleached samples are reported in light gray and those on untreated ones are reported in dark gray.

Calcein loading was calculated with respect to the calcium content measured on pristine precipitates and bleached ones. Calcite crystals entrap significantly higher quantity of calcein as compared to aragonite ones ($p \leq 0.05$, figure 1). Interestingly, bleaching does not significantly change the content of calcein in calcite crystals ($p > 0.05$), while in aragonite crystals it significantly decreases after bleaching ($p \leq 0.05$), implying that the quantity of surface adsorbed dye is meaningful only for aragonite. Indeed, it has been reported that aragonite has a surface area almost ten times higher than that of calcite, justifying a higher adsorption efficiency for mass unit of $CaCO_3$.[27]

The experimental data indicate that the amount of dye entrapped within calcium carbonate increases increasing its concentration solution. On the other side, the percentage of dye removed from the solution during the crystallization process decreases with its concentration. This observation indicates that 40 µM, the concentration used in *in vivo* experiments, does not allow a maximum of loading into calcium carbonate in the experiments carried out *in vitro*. In fact, increasing the concentration to 400 µM results in the increase of the loading of almost ten times. Thus, if an organism accumulates calcein at the mineral growing sites, a higher entrapment can be achieved, with an enhancement of the fluorescence signal.

The spatial distribution of calcein entrapped within the crystals was investigated by confocal microscopy. This analysis was carried out on the samples chemically bleached prior to the imaging to exclude any fluorescence contribute from the dye adsorbed onto crystal surfaces.

Confocal images of calcite crystals grown in the presence of calcein (figure 2a-c) show an inhomogeneous distribution of the dye within the crystals. The dye localizes preferentially on rhombohedral faces, which in the spherulitic calcite (figure 2c) are more extended in the external layers. In addition to this, it cannot be excluded that the signal from the center of the spherulitic crystals is not detected, since the thickness of the sample and the geometry of the microscope do not allow the laser to reach the top layers (see figure SI1). Confocal images of aragonite crystals grown in the presence of calcein (figure 2d-f) show a homogeneous distribution of the dye inside the crystals. Since aragonite particles, as well as those of magnesium calcite, are composed of packed acicular crystals [28], it can be assumed that the dye is mainly entrapped among the crystalline domains, and only partly is embedded within the lattice. This observation is also confirmed by the results on dye loading, which show a significant reduction of loading after bleaching.

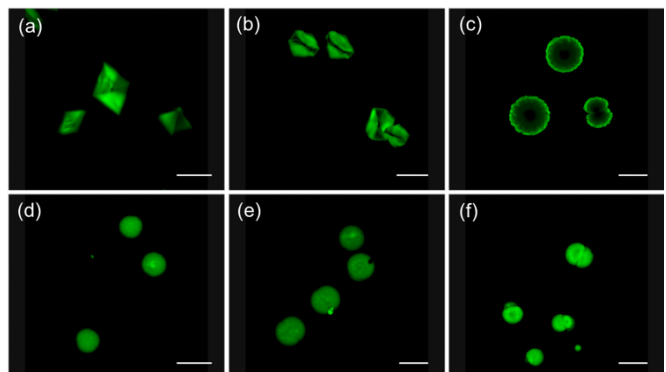

**Figure 2.** Confocal micrographs of calcite crystals grown in the presence of (a) 4 µM, (b) 40 µM and (c) 400 µM calcein and aragonite crystals grown in the presence of (d) 4 µM, (e) 40 µM and (f) 400 µM calcein. Scalebar is 50 µm.

Rietveld refinement applied to the high resolution X-ray powder diffraction data allowed quantifying the $CaCO_3$ polymorphic distribution and the lattice distortions (see table SI4) induced by calcein interaction with calcium carbonate crystals.[18,29] The crystalline phase analysis shows that calcein does not affect the precipitation of calcite, which always occurs regardless its concentration. Interestingly, in the presence of calcein pure calcite is obtained, while in the absence of the additive traces of aragonite and vaterite are present (see figure SI3). The scenario is different in the precipitation conditions used for aragonite. The presence of calcein promotes precipitation of magnesium calcite on the expense of aragonite and when 400 µM calcein is present in the crystallization solution only 4 wt.% of aragonite co-precipitate (see figure SI4). This trend could be related to the higher complexation constant of calcein for $Ca^{2+}$ than for $Mg^{2+}$ [30] that, increasing the concentration of free $Mg^{2+}$ ions with respect to $Ca^{2+}$, affects $CaCO_3$ supersaturation of the precipitating solution and the thermodynamic stability of magnesium calcite and aragonite.[31] However, it has been reported that the presence of calcein does not impact the incorporation of $Mg^{2+}$ into biologically and inorganically precipitated calcium carbonate.[32]

**Figure 3.** (a) The (104) calcite diffraction peak for control calcite (blue) and calcite grown with 400 µM calcein (red) and (b) the (111) aragonite diffraction peak for control aragonite (blue) and aragonite grown with 400 µM calcein (red). Wavelength converted from 0.4959 Å to 1.5406 Å.

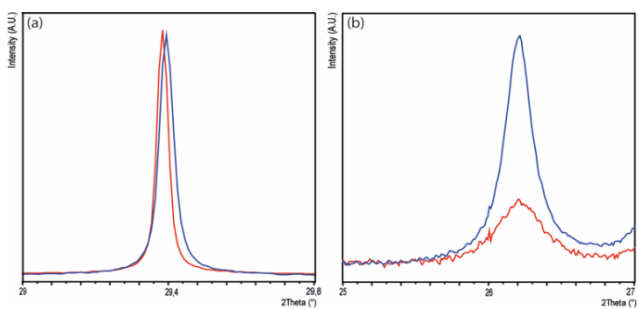

The presence of calcein also induces a change in the lattice parameters of calcite (see table SI5). This change is visualized in figure 3a where the (104) diffraction peak of calcite shifts to a lower 2Θ value when it is grown in the presence of 400 µM of calcein. This is mainly related to an expansion of the c-axis (Table SI5) associable to the entrapping of the additive. No significant shift of aragonite peaks is observed (figure 3b), indicating that calcein is mainly localized among the crystalline units of aragonite, as suggested by the bleaching results and confocal microscopy experiments and the non-significant variation of the lattice parameters (table SI6). In the magnesium calcite crystals, the lattice distortions are due to a combined opposed effect of calcein and $Mg^{2+}$ ions, with the former expanding the c-axis, as observed for calcite and the latter decreasing the a-axis (table SI6).

SEM images show that the dye modifies calcite crystal shape and morphology (figure 4a-d). When grown in the presence of 4 or 40 µM calcein, the crystals show only {*10.4*} rhombohedral faces (figure 4b) or demonstrate additional crystalline {*hk.l*} faces almost parallel to the c-axis, respectively. The presence of 400 µM calcein in the crystallization solution leads to the formation of spherulitic aggregates exposing small {*10.4*} faces. A similar trend in the evolution of the shape and morphology with the increase of the concentration of an additive has been observed in the presence of block copolymer poly(ethylene glycol)-block-poly(methacrylic acid), [33] which similarly to calcein contains carboxylate functional groups. Aragonite crystals precipitated in the absence of additives (figure 4e) present a dumbbell shape, composed of needles with a triangular section. Crystals grown in the presence of 4 µM or 40 µM calcein (figure 4 f-g) appear elongated and show {*hk0*} faces, which dimensions decrease with the increase of the concentration of the additive. The mass ratio of magnesium calcite increases at higher calcein concentration and, when crystals are grown in the presence of 400 µM calcein (figure 4 h), they are composed of small particles aggregated in a dumbbell shape in which the characteristic stepped {01.1} faces are observable. [34]

In conclusion, we studied the effect of calcein, a dye commonly used to mark the growth of calcifying organisms, on the growth of calcite and aragonite crystals *in vitro*. These data reveal a strong influence of calcein on the precipitation of calcium carbonate, affecting both polymorphic distribution and crystal morphology. Calcein is loaded into the structure of both polymorphs, localizing within the lattice preferentially along (104) planes in calcite crystals and adsorbing among crystalline domains in aragonite. These *in vitro* observations do not find a documented correspondence in *in vivo* systems, for which no effect of calcein on the calcification process have been reported so far. Thus, it appears evident that the organisms' capability to control the deposition of the mineral phases can overcome the potential interference caused by calcein.

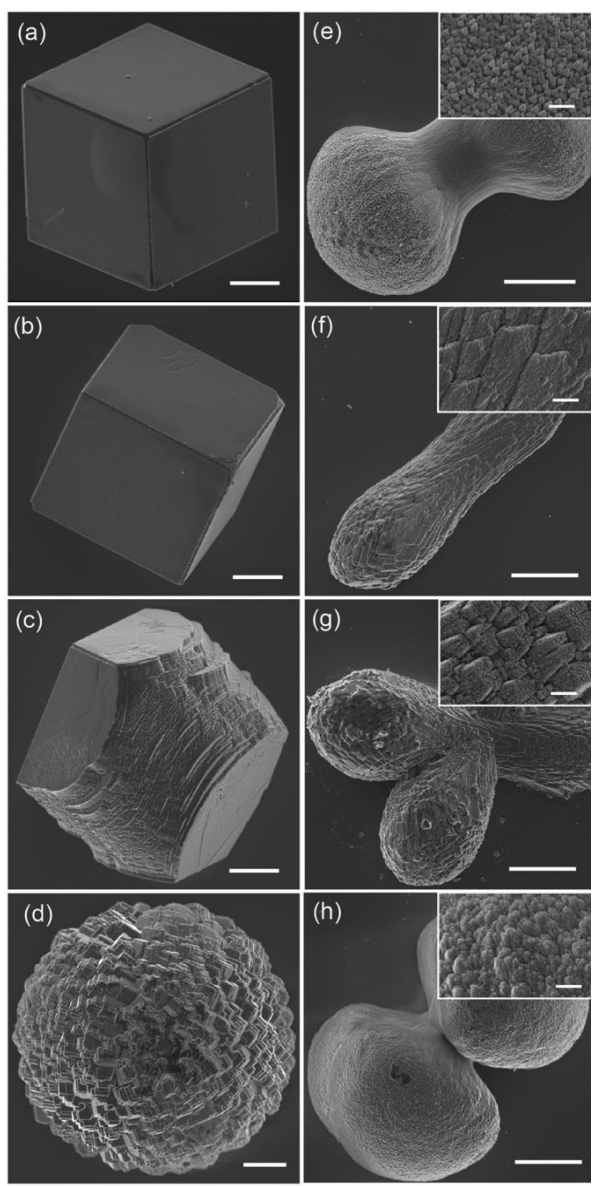

**Figure 4.** SEM images of crystals grown in the presence of 10 mM $Ca^{2+}$ (a) without additives and in the presence of (b) 4 µM, (c) 40 µM and (d) 400 µM calcein and crystals grown in the presence of 10 mM $Ca^{2+}$ and 40 mM $Mg^{2+}$ (e) without additives and in the presence of (f) 4 µM, (g) 40 µM and (h) 400 µM. Scalebar is 10 µm in the main picture and 1 µm in the inset. Images are representative of the whole crystal population (see figure SI1).

## Conflicts of interest

There are no conflicts to declare.